\documentclass{icrc}

\usepackage{times}
\usepackage{graphicx} 

\begin{document}

\title{Flux of upward high-energy muons at the multi-component
primary energy spectrum}
\author[1]{S.V. Ter-Antonyan}
\affil[1]{Yerevan Physics Institute, 2 Alikhanian
Br. Str., 375036 Yerevan, Armenia}
\author[2]{P.L. Biermann}
\affil[2]{Max-Planck-Institute f\"ur Radioastronomie
Auf dem H\"ugel 69, D-53121 Bonn, Germany}

\correspondence{samvel@jerewan1.yerphi.am}

\firstpage{1}
\pubyear{2001}


\maketitle
\begin{abstract}
The atmospheric neutrino-induced upward muon flux are
calculated by using the multi-component primary energy spectrum,
CORSIKA EAS simulation code for the reproduction of the
atmospheric neutrino spectra and improved parton model for
charged-current cross sections. The results are obtained at
$10^2-10^6$ GeV muon energy range and $0-89^0$ zenith angular
range.
\end{abstract}

\section{Introduction}
Neutrino-induced upward muon flux are both the direct transmitters
of information about neutrino flux from energetic astrophysical
sources (binary stars, AGN) and direct posterity of atmospheric very
high energy neutrino. These problems are being investigated
in many modern
experiments (see review \citep{GHS}) and as a rule, the upward muon
flux of atmospheric origin is considered in the capacity of the
background flux.\\
Here, on the basis of multi-component model of primary cosmic ray
\citep{PeBi}, QGSJET model of high energy $A-A_{Air}$ interactions
\citep{QGS} and parton model predictions of charged-current cross
sections \citep{QRW} we
evaluated the expected background upward muon energy spectra at
different zenith angles and studied 
($E_{\mu}-E_{\nu,\overline{\nu}}$) and ($E_{\mu}-E_{N}$) 
correlations.

\section{Method of calculations} 
Upward muon differential 
energy spectra close to Earth surface at different zenith angles
($\theta$) can be represented as follows:
\begin{eqnarray} 
\frac{\partial F(\theta)}{\partial E_{\mu}} & = & 
\sum_{A} \int_{E_{\min}}^{E_{max}} 
\frac{d\Im} {dE_{A}}dE_{A}
\sum_{\xi\equiv\nu_{\mu},\overline{\nu}_{\mu}}
\int_{E_{\mu}}^{E_{A}} 
\frac{\partial f}{\partial E_{\xi}}\cdot{} 
                              \nonumber \\
 &  & {} \cdot
dE_{\xi}
\int_{t_{\min}}^{t_{\max}} 
\frac{\partial G_{\xi}}{\partial
E_{\mu}} \frac{\partial W}{\partial t}dt 
\end{eqnarray} 
where:\\
$d\Im/E_{A}$ - is the primary energy spectrum of the nucleus
$A$;\\
$\partial f(E_{A},A,\theta)/\partial E_{\xi}$ is the atmospheric
neutrino ($\xi\equiv\nu_{\mu}$) and anti-neutrino
($\xi\equiv\overline{\nu}_{\mu}$) energy spectrum at the primary
nucleus $A$ with $E_{A}$ energy and $\theta$ zenith angle;\\
$\partial G_{\xi}(t,E_{\xi})/\partial E_{\mu}$ is the
neutrino-induced muon energy spectrum at observation level (close
to the Earth surface) at
 \begin{figure}[t]
 \includegraphics[width=8.3cm]{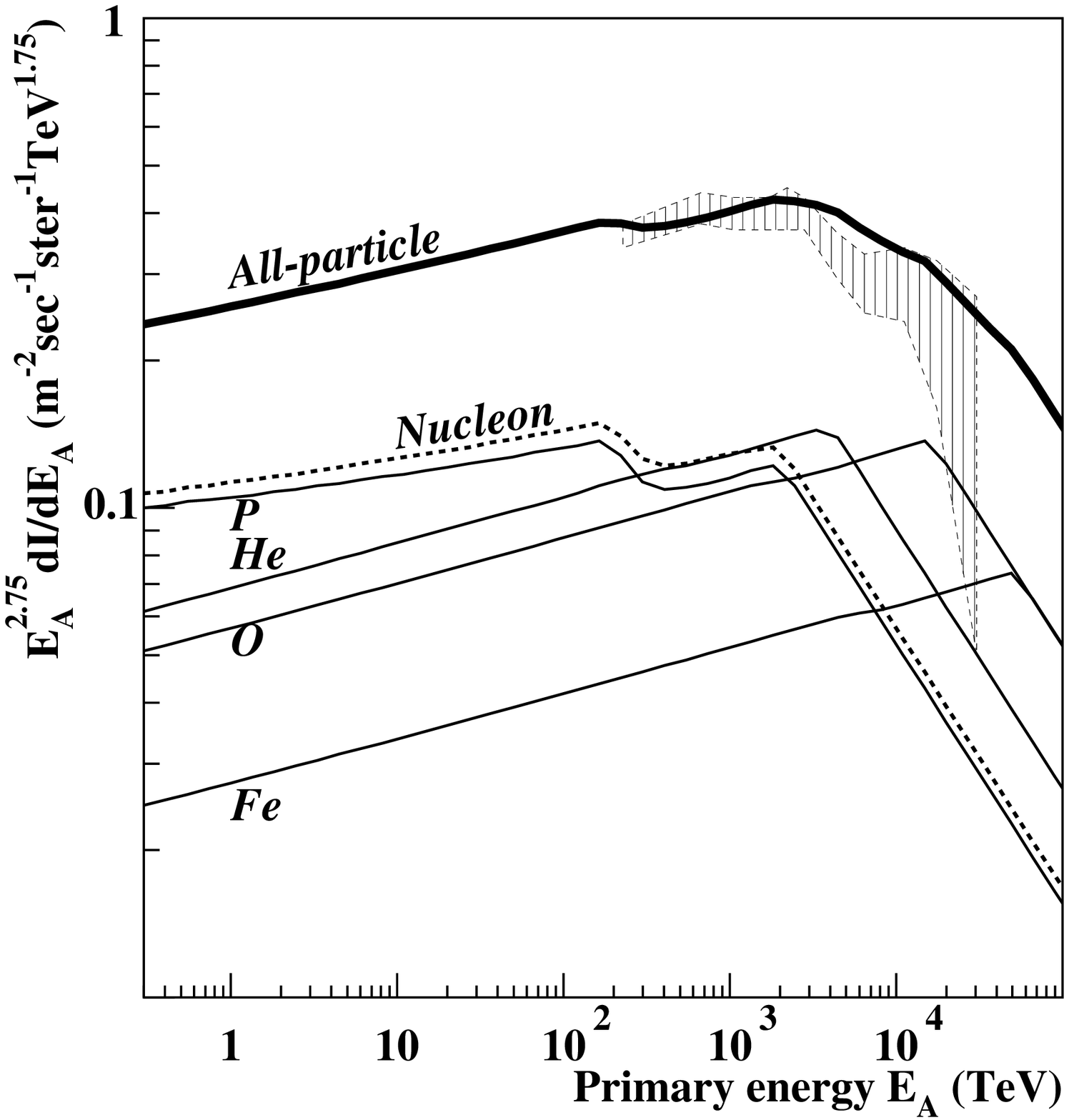} %
 \caption{Primary energy spectra according to 2-component origin of
cosmic ray. The hatch area contains \citep{TH}, DICE
\citep{DICE}, and CASA-BLANCA \citep{CB} data.}
\end{figure}
the given neutrino energy $E_{\nu,\overline{\nu}}$ and 
depth $t$ of muon production in the Earth;\\ 
$\partial W(E_{\xi},E_{\mu})/\partial t$
is a probability of charged-current interaction at $t$ depth of  
the Earth matter.\\ 
Energy spectra of primary nuclei ($A\equiv1-59$)  
according to the multi-component model of primary cosmic ray origin
\citep{PeBi} are presented in 2-component form (see
Fig.1):
\begin{figure*}[t]
 \includegraphics[width=17.0cm,height=12cm]{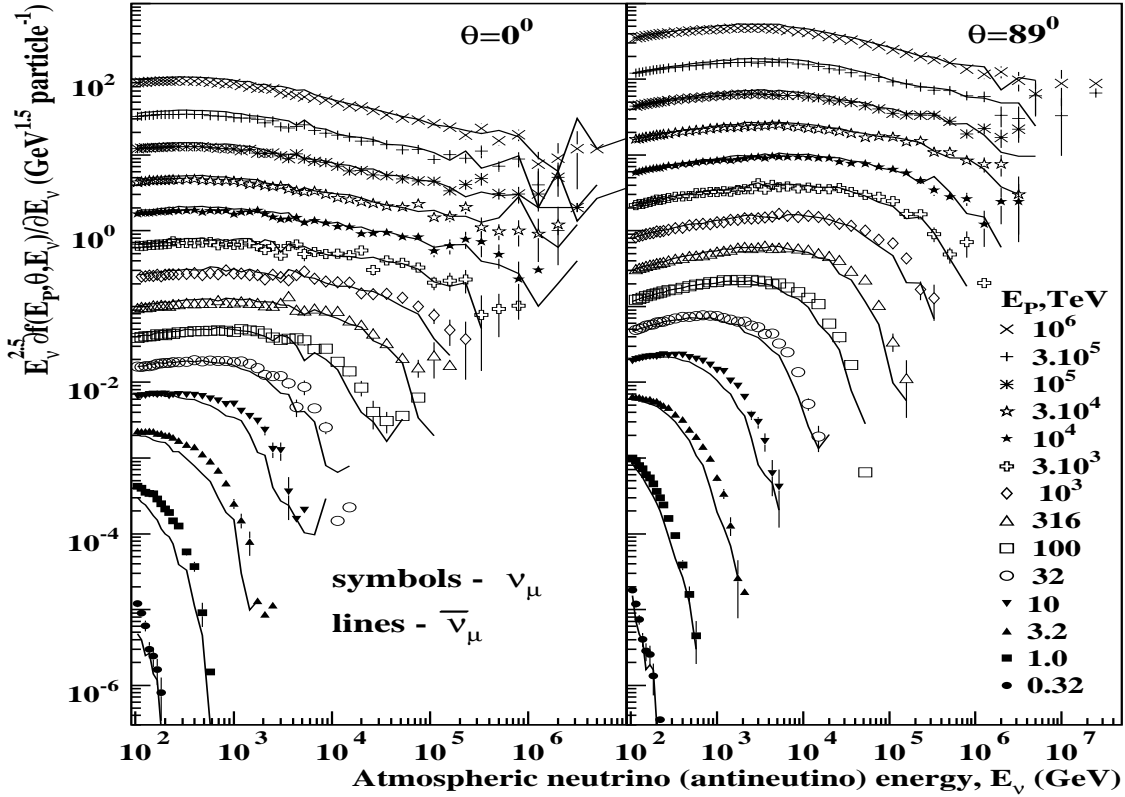} %
 \caption{Atmospheric neutrino (symbols) and anti-neutrino 
(lines) energy spectra at different primary proton energy ($E_p, TeV$) at
zenith angles $\theta=0^0$ (left) and $\theta=89^0$ (right).}
 \end{figure*}
\begin{equation}
\frac{\partial \Im} {\partial E_{A}}= 
\Phi_{A}\Big(
\delta_{A,1}\frac{d\Im_{1}}{dE_{A}}+
\delta_{A,2}\frac{d\Im_{2}}{dE_{A}}
\Big) 
\end{equation} 
where the first component (ISM) is
derived from the explosions of normal supernova into an
interstellar medium with expected rigidity-dependent power law
spectra 
\begin{equation} 
\frac{d\Im_{1}}{dE_{A}} = \left\{
\begin{array}{l@{\quad:\quad}l} 
E_{A}^{-\gamma_{1}} &
E_{A} < E_{ISM} \\
0 & E_A>E_{ISM} 
\end{array} \right.
\end{equation} 
and the second component (SW) is a result of the explosions of
stars into their former stellar winds with expected
rigidity-dependent power law spectra 
\begin{equation}
\frac{d\Im_{2}}{dE_{A}} = 
\left\{ \begin{array}{l@{\quad:\quad}l}
E_{A}^{-\gamma_{2}} & E_{A} < E_{SW} \\
E_{SW}^{-\gamma_{2}}(E_{A}/E_{SW})^{-\gamma_{3}} &
E_A>E_{SW} \end{array} \right. 
\end{equation} 
where $\Phi_{A}$ is a scale factor ($E$ in TeV units) from
approximations \citep{BWPB},\\ 
$E_{ISM}=R_{ISM}\cdot Z$ and $E_{SW}=R_{SW}\cdot Z$ are the
corresponding rigidity-dependent critical energies \citep{PeBi}
and $Z$ is the charge of nucleus $A$.\\
The values of spectral
parameters from (2-4) are:\\ 
$\gamma_1=2.78$, $\gamma_2=2.65$, $\gamma_3=3.28$,
$R_{ISM}\simeq200$ TV, $R_{SW}\simeq 2000$ TV and the 
fraction of each component: \\
$\delta_{A,1}=1-\delta_{A,2}$, 
$\delta_{A,2}=(2ZR_{ISM})^{(\gamma_{2}-\gamma_{A,0})}$ at
$\gamma_{A=1,0}=2.75$ and $\gamma_{A>1,0}=2.66$ \citep{BWPB} is
obtained by normalization of (2-4) with approximation of balloon and
satellite data \citep{BWPB} at $E_{A}=1$ TeV and $\chi^2$
minimization of KASCADE \citep{KAS} and ANI \citep{ANI} EAS size
spectra with the expected ones using 2-component representation 
(2-4) \citep{BONNY}.\\
Values of primary energy limits in the expression (1) 
are chosen $E_{min}=2E_{\mu}$ and $E_{max}=10^9$ GeV. 
The expected primary energy spectra for
different nuclei, all-particle and nucleon energy spectra
by 2-component model are shown in the Fig.~1. A contribution
of the third primary extra-galactic component is ignored.\\ 

Atmospheric neutrino
(anti-neutrino) energy spectra\\ 
$\partial f(E_{A}, A, \theta)/\partial E_{\xi}$ at given six $A$
group ($A=1$, $4$, $12$, $16$, $28$,
$56$) and $E_{A}=3.16\cdot10^2,10^3,\dots10^9$ GeV, 
$\theta=0,30,60,89^0$ parameters of primary nuclei were calculated
in the framework of QGSJET interaction model \citep{QGS} by
CORSIKA562 EAS
simulation code \citep{COR}. Intermediate values of tabulated
spectral
data are evaluated by interpolation with preliminary
linearizations. It should be noted that the neutrino energy spectra
depend only on primary nucleon energy ($E_{A}/A$) with accuracy 
$<5\%$ in $A\equiv1-56$ range.\\ 
The atmospheric neutrino
(anti-neutrino) energy spectra at different primary proton
energies ($E_{p}=3.16\cdot10^2-10^9$ GeV) and zenith angles
($\theta=0^0, 89^0$) are presented in Fig.~2.\\
Neutrino-induced muon
energy spectra are determined as 
 \begin{figure}[t]
 \includegraphics[width=8.3cm]{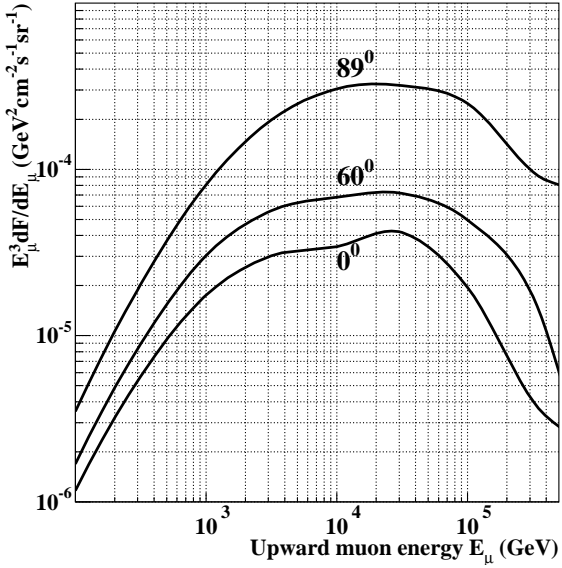} %
 \caption{Atmospheric neutrino-induced 
upward muon differential energy spectra.}
 \end{figure}
\begin{equation} 
\frac{\partial
G_{\xi}}{\partial E_{\mu}}= \frac{1}{\sigma_{\xi}(y<y_{\max})}
\frac{d\sigma_{\xi}}{dy} \frac{e^{bt}}{E_{\xi}} 
\end{equation}
where 
\begin{equation} \frac{d\sigma_{\xi}}{dy}=
\int_{x_{\min}}^{1} \frac{\partial^{2}\sigma_{\xi}}{\partial
x\partial y}dx\;\;, 
\end{equation} 
\begin{equation}
\sigma_{\xi}(y<y_{\max})= \int_{0}^{y_{\max}} \frac{\partial
\sigma_{\xi}}{\partial y}dy 
\end{equation} 
and
$\partial^{2}\sigma_{\xi}/\partial x\partial y$ are corresponding
inclusive charged-current cross sections for the reaction
$\nu_{\mu}(\overline{\nu}_{\mu})  + N\rightarrow
\mu^{+}(\mu^{-})+$ anything. 
Here $\xi\equiv\nu_{\mu},\overline{\nu}_{\mu}$, 
$x=Q^2/2M\nu$ and $y=\nu/E_{\xi}$ are the scaling variables,
$Q^2$ is an invariant
momentum transfer between an incident neutrino and an outgoing muon,
$\nu=E_{\xi}-E_{\mu}^*$ is the energy loss in a laboratory
frame, $M$ is the nucleon mass, $x_{\min}=5\cdot10^{-6}$,
\begin{equation}
y_{max}=1-E_{\mu}^{*}/E_{\xi},
\end{equation}
\begin{equation}
E_{\mu}^{*}=(a+bE_{\mu})\exp[b(t_{\max}-t)]/b-a/b
\end{equation}
is the muon
energy in production point $t$, parameters
$a=0.002$ MeV$\cdot$cm$^2$/g and
$b=4\cdot10^{-6}$ cm$^2$/g \citep{GHS} determine muon
energy losses: $-dE_{\mu}/dt=a+bE_{\mu}$.\\ 
The probability of charged-current interaction in expression (1) at 
given depth $t$ of the Earth matter has a form:
\begin{equation} 
\frac{\partial W}{\partial t}=
\sigma_{\xi}N_{A}exp(-\sigma_{\xi}N_{A}t) 
\end{equation} 
where $\sigma_{\xi}\equiv
\sigma_{\nu,\overline{\nu}}(y<y_{max})$ is a total
charged-current cross section (7) and 
$N_{A}$ is the Avogadro number. \\
The depth limits of effective ($y<y_{max}$)
charged-current interactions in the Earth matter ($t$) are
\begin{equation}
t_{\min}=t_{max}-L(E_{\xi},E_{\mu}),
\end{equation}
\begin{equation}
t_{\max}(\theta)=\int_{0}^{z_{\max}}\rho(r(z))dz. 
\end{equation}
Here
$\rho(g/cm^2)=13.8-1.73\cdot10^{-8}r$(cm) is our approximation
of the Earth density at the radius $r$, $z_{\max}=2R\cos{\theta}$,
$R=6.371\cdot10^8$cm is the Earth radius,
\begin{equation}
L=(1/b)\ln[(a+bE_{\xi})/(a+bE_{\mu})] 
\end{equation}
is a rock
thickness for muon detection with energy $E_{\mu}$ on the Earth
surface at $y=0$.

\section{Results} 
 \begin{figure}[t]
 \includegraphics[width=8.3cm]{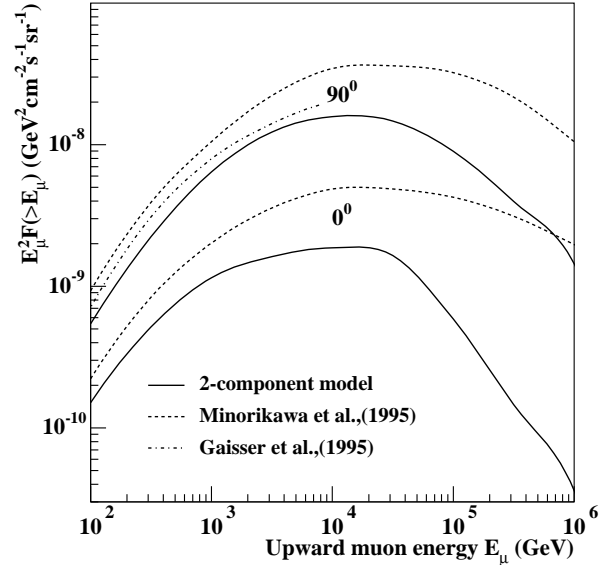} %
 \caption{Integral upward muon energy spectra. The data \citep{GHS}
obtained at $-0.3<\cos\theta<0.3$.}
 \end{figure}
Calculations of upward muon
energy spectra according to expression (1) are performed using
numerical integration. The charged-current cross section for muon
production by neutrino was taken from
the renormalization-group-improved parton model
\citep{QRW} with quark structure functions for $x>10^{-4}$ from
\citep{EHLQ}. We limited the range of $x<10^{-4}$ by
$x_{min}=5\cdot10^{-6}$ because the check of valence
quark structure functions at very small $x$ values by  
quark number sum rules
\begin{equation}
\int dx\;u_v(x,Q^2)=2
\end{equation}
\begin{equation}
\int dx\;d_{v}(x,Q^2)=1
\end{equation}
showed that the obtained errors
of sum rules became more than $1-2\%$ at $x<x_{min}$.\\
\begin{figure}[t]
 \includegraphics[width=8.3cm]{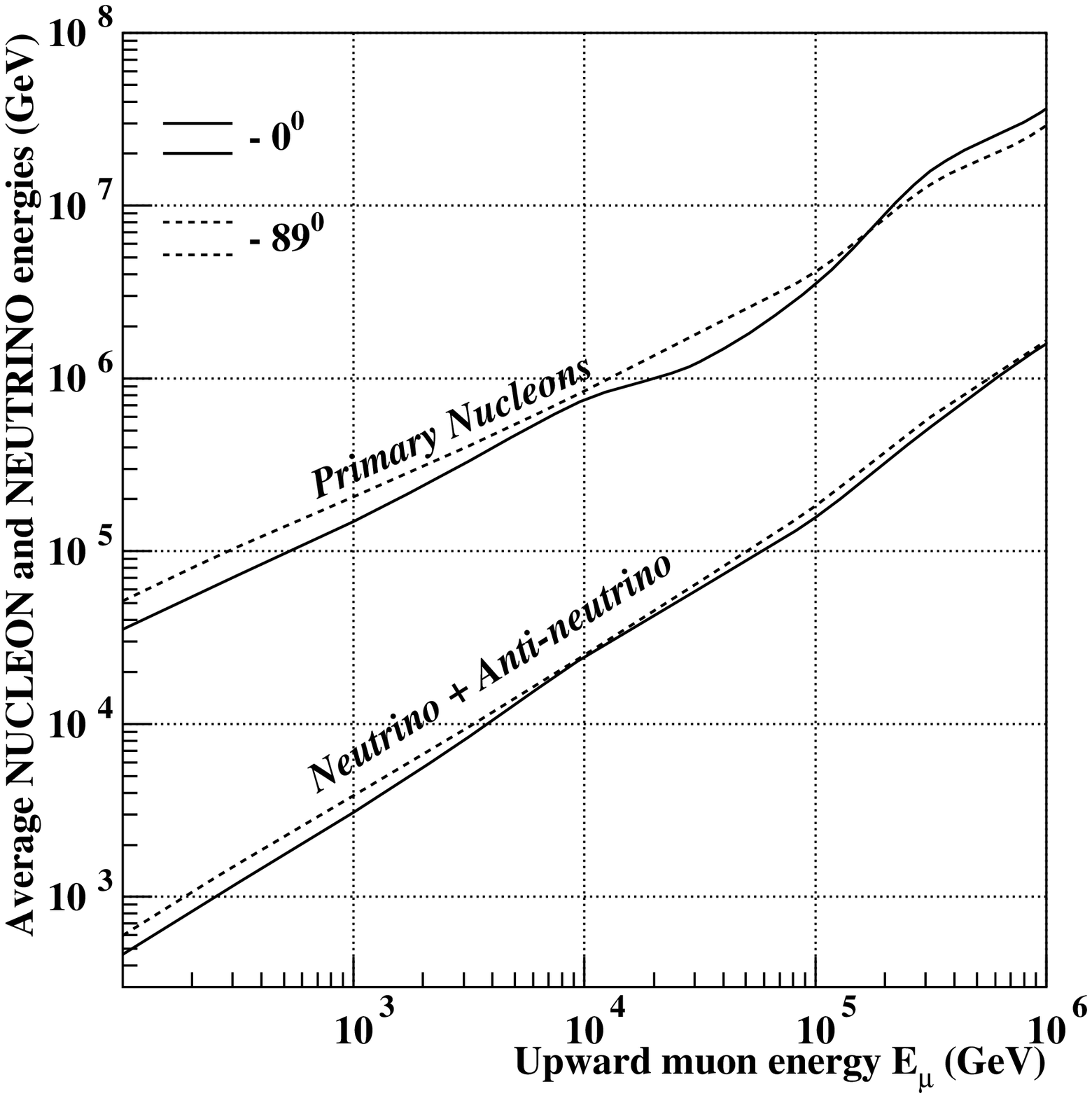} %
 \caption{($E_{\mu}-<E_{N}>$) and
($E_{\mu}-<E_{\nu,\overline{\nu}}>$) correlations.}
 \end{figure}
Upward muon differential  energy spectra at different 
zenith angles are presented in Fig.~3. An accuracy of
multidimensional
integration (1) was less than $3\%$.
Corresponding upward muon integral energy spectra (with accuracy
$4-5\%$) in comparison with
the same calculations \citep{Rome} are presented in Fig.~4.

\newpage 
Correlations of muon energy at observation level
(near the Earth surface) 
with primary nucleon and parent-neutrino energies are shown
in Fig.~5. Some irregularities of presented data at very high
energies ($E_{\mu}>3\div5\cdot10^5$ GeV)
in Fig.~3-5 are explained by the insufficient simulation
sampling for atmospheric neutrino energy spectra (fig.~2) at 
corresponding energies and accuracies of multi-dimensional 
integrations.
\begin{acknowledgements}
We thank Johannes Knapp and Dieter Heck for providing the CORSIKA
code. The work has been partly supported by the research grant N
00-784
of the Armenian government,
NATO NIG-975436 and CLG-975959 grants and ISTC A216 grant.
\end{acknowledgements}

\end{document}